\documentclass[12pt] {article}

\usepackage{amssymb,amsmath,latexsym, mathrsfs}

\renewcommand{\appendix}{\setcounter{section}{0}\renewcommand{\thesection}{}
\renewcommand{\thesubsection}{\Alph{subsection}.}
\section{Appendix}
}
\def\Appendix#1{
 			\setcounter{equation}{0}
 			\renewcommand{\theequation}{\thesubsection\arabic{equation}}
 			\subsection{#1}
 			}

\newcommand{\beq}[1]{  \\ {\tiny ({#1})}   \begin{equation} \label{#1} }
\newcommand{\beqa}[1]{ \\ {\tiny ({#1})}   \begin{eqnarray} \label{#1} }
\renewcommand{\beq}[1]{  \begin{equation} \label{#1} }
\renewcommand{\beqa}[1]{\begin{eqnarray} \label{#1} }

\newcommand{\eeq}{\end{equation}}\newcommand{\eeqa}{\end{eqnarray}}
\newcommand{\rf}[1]{(\ref{#1})}
   
\def\tr{\operatorname{tr}}
\def\skew{\operatorname{\bf skew}}
\def\diag{\operatorname{diag}}
\def\tens#1{\mathbb{\,#1}}
\def\bd#1{\mbox{\boldmath$\displaystyle\mathbf{#1}$}}
\def\doublespacing{\baselineskip=18pt}
\usepackage{color}
\newif\ifpdf
\ifx\pdfoutput\undefined
      \pdffalse
\else
      \pdftrue
\fi
 
\ifpdf
    \pdfcatalog { /PageMode (/UseNone)
                  /OpenAction (fitbh)
    }
 
  \usepackage[pdftex]{graphicx}
  \usepackage[pdftex]{hyperref}
  \hypersetup{
   pdftitle={Fedorov's problem},
    pdfsubject={acoustics},
    pdfkeywords={elasticity,waves,seismics,ultrasonics},
    pdfauthor={Andrew N. Norris,
               <norris@rutgers.edu>},
    pdfpagemode={UseOutlines},
    bookmarksopen=true,
    colorlinks=true,
    urlcolor=rltblue,
    filecolor=rltgreen,
    linkcolor=rltred,
    citecolor=blue,
    pagecolor=red,
    urlcolor=cyan
  }
  \definecolor{rltred}{rgb}{0.75,0,0}
  \definecolor{rltgreen}{rgb}{0,0.5,0}
  \definecolor{rltblue}{rgb}{0,0,0.75}
\else
  \usepackage{graphicx}
\fi

\if@mathematic
   \def\bd#1{\ensuremath{\mathchoice
                     {\mbox{\boldmath$\displaystyle\mathbf{#1}$}}
                     {\mbox{\boldmath$\textstyle\mathbf{#1}$}}
                     {\mbox{\boldmath$\scriptstyle\mathbf{#1}$}}
                     {\mbox{\boldmath$\scriptscriptstyle\mathbf{#1}$}}}}
\else
   \def\bd#1{\ensuremath{\mathchoice
                     {\mbox{\boldmath$\displaystyle#1$}}
                     {\mbox{\boldmath$\textstyle#1$}}
                     {\mbox{\boldmath$\scriptstyle#1$}}
                     {\mbox{\boldmath$\scriptscriptstyle#1$}}}}
\begin{document} %%%%%%%%%%%%%%%%%%%%%%%%%%%%%%%%%%%%%%%%%%%%%%%%%%%%%%%%%%%%%%%%%%%%%%%%%
\doublespacing %\singlespacing%

\title{
%\textcolor{blue}
{Elastic moduli approximation of higher symmetry for  the acoustical properties  of an anisotropic material}
}
\author{Andrew N. Norris\\ 
\\ Rutgers University\\ Department of Mechanical and Aerospace Engineering\\ 98 Brett Road, Piscataway, NJ  08854-8058
}
\date{}
\maketitle

\begin{abstract}

The issue of how to define and determine an optimal  acoustical fit to a set of anisotropic elastic constants is addressed.   The optimal moduli are defined as those which  minimize the mean squared  difference in the acoustical tensors between the given moduli and all possible moduli of a chosen higher material symmetry.  The solution is shown  to be identical to  minimizing a Euclidean distance function, or equivalently,  projecting the tensor of elastic stiffness  onto the appropriate symmetry.   This has implications for how to best select anisotropic constants to acoustically model complex materials.  

\end{abstract}

~ \\
Running title: Elastic moduli approximation \\ ~ \\
Submitted: November 5, 2005 \\ ~ \\
Contact:  norris@rutgers.edu 

\newpage 
\section{Introduction}

The theory of wave propagation in anisotropic elastic solids  was historically developed for homogeneous crystals with {\it a priori} known symmetry \cite{Musgrave}.  The number of elastic moduli are then well defined.   
However,  the  symmetry is not always apparent  in non-crystalline and generally complex materials for which measurements can yield as many as 21  anisotropic elastic coefficients.  Whether one has some idea of the underlying material symmetry or simply prefers to deal with fewer parameters,  the question arises of how to best fit the given  set of  elastic moduli to, for instance,  a transversely isotropic material model.  This issue occurs in acoustical measurements of composites composites \cite{fgm}, and in geophysical applications \cite{Browaeys04} where laboratory measurements might yield 21 moduli, but seismic modeling requires a higher symmetry, such as transverse isotropy.   The purpose  of this paper is to provide a simple but unambiguous means to find the reduced set of anisotropic elastic constants that are in a certain sense the best  acoustical fit to the given moduli.   The optimal material minimizes the mean square difference in the slowness surfaces of the given moduli and of all possible sets of elastic constants of the chosen symmetry. 

The prevailing approach to finding a reduced set of moduli  does not invoke acoustical properties, but views the moduli as elements in a vector space which are projected  onto the  higher elastic material symmetry.   This is achieved by defining a 
 Euclidean norm for the moduli $\tens C$ according to $\| \tens C \|^2 = C_{ijkl}C_{ijkl}$ where\footnote{Lower case Latin suffices take on the values 1, 2, and 3, and
the summation convention on repeated indices is assumed.} $C_{ijkl}$ are the elements of the stiffness tensor. This   provides a natural definition for distance, from which  one can find the elastic tensor of a given symmetry nearest to an anisotropic elastic tensor, or equivalently, define a projection appropriate to the higher symmetry.  Gazis et al. \cite{Gazis63} outline the procedure in terms of fourth order tensors, while  more recently Browaeys and Chevrot  \cite{Browaeys04} provide projection matrices for $\tens C$ expressed as a 21-dimensional vector.  Helbig \cite{Helbig95} provides a useful overview of the problem from a geophysical perspective,  Cavallini \cite{Cavallini99} examines isotropic projection specifically, and Gangi \cite{Gangi00} gives formulas for several other symmetries.   One drawback of the approach is that while it minimizes the Euclidean distance  between the original and projected stiffness tensors, it does not provide the analogous  closest {\em compliance} (the inverse of stiffness).  In this sense the stiffness projection method, while simple and attractive, is unsatisfactory because it is not invariant under inversion.   
 Alternative procedures based on non-Euclidean norms such as the Riemannian  \cite{Moakher04} or log-Euclidean \cite{Arsigny:MICCAI:05} metrics do not have this deficiency, and, in principle, provide a unique ``projection" regardless of whether one uses the stiffness or the compliance tensor. 

An apparently quite different approach is to try to find higher symmetry moduli  which in some way better approximate the acoustic properties of the given moduli.  Thus, Fedorov \cite{fed} considered the question of what elastically {\em isotropic} material is the best acoustic fit to a given set of 
%$n,\, 2 < n\le 21$, 
anisotropic moduli.  He defined best fit to mean the effective bulk and shear moduli $\{\kappa,\, \mu\}$ which  minimize the mean square  difference between the slowness surfaces of the  original anisotropic material and the isotropic material characterized by $\{\kappa,\, \mu\}$ (density is unaffected).  Fedorov obtained  explicit expressions for the  moduli, eq. (26.19) of \cite{fed}, or in the present notation
\beq{ie}
\kappa = \frac19\, C_{iijj},\qquad 
\mu = \frac1{10}\, C_{ijij} - \frac1{30}\, C_{iijj}\, . 
\eeq

Fedorov's procedure for finding a suitable set of higher symmetry moduli is physically appealing, especially as it seeks to approximate acoustical properties.  Also,  as Fedorov and others \cite{fed,Gazis63,Browaeys04} have shown,  $\kappa$ and $\mu$ are precisely the isotropic moduli found  by the stiffness projection method. 
However, Fedorov only considered effective isotropic moduli and it does not appear that anyone has attempted to generalize his method to symmetries other than isotropic.   
The purpose of this paper is to solve what may be termed the generalized Fedorov problem.  The solution is found for the acoustically best fitting moduli of arbitrary symmetry, which is of higher symmetry than the given moduli but lower than isotropy.  The central result is that the solution of the generalized Fedorov problem possesses  the same crucial property as the solution obtained by Fedorov \cite{fed}, that is, the best fit moduli are identical to those obtained by the stiffness projection method. 
This result provides a strong physical and acoustical  basis  for using the Euclidean projection scheme that has been absent until now.  

We begin in Section \ref{sec2} with the definition of  elastic tensors and associated notation.  The generalized Fedorov problem is introduced and solved in Section \ref{sec3}.  Examples are given  in    Section \ref{ex}.

\section{Preliminaries}\label{sec2}

 \subsection{The elasticity   tensor and related notation}

The solution of the generalized Fedorov problem is most easily accomplished using tensors, which are reviewed here  along with some relevant notation.  
 Boldface lower case Latin quantities  indicate 3-dimensional vectors,  such as the orthonormal basis $\{ {\bd e}_i, \, i=1,2,3\}$,   ``ghostface" symbols such as  $\tens C$ indicate fourth order elasticity tensors, and  boldface capitals, e.g. ${\bf A}$, are second order symmetric tensors, with some exceptions.  Components  defined relative to the  basis  vectors allow us to represent arbitrary tensors in terms of the fundamental tensors formed from the basis vectors, thus,  
  ${\bf A} = A_{ij} \, {\bd e}_i \otimes {\bd e}_j$,  
  ${\tens C} = C_{ijkl}\,  {\bd e}_i \otimes{\bd e}_j \otimes{\bd e}_k \otimes{\bd e}_l$, where 
  $\otimes $ is the tensor product.    
The   identity tensors of second and   fourth order are $\bd I$ and  ${\tens I}$, respectively, and we will also use the  fourth order tensor ${\tens J}$.  These have components
\beq{comp}
I_{ij} = \delta_{ij}, \qquad  
I_{ijkl} = \frac12
(\delta_{ik}\delta_{jl} +\delta_{il}\delta_{jk} ), \qquad 
J_{ijkl} = \frac13 \delta_{ij}\delta_{kl}  \, . 
 \eeq
Isotropic elasticity  tensors can be expressed as a linear combination of ${\tens I}$ and ${\tens J}$. Using Lam\'e moduli   $\mu$ and $\lambda = \kappa - \tfrac23 \mu$,   for instance,   
\beq{607}
{\tens C} = 2\mu {\tens I} + 3\lambda {\tens J}\, . 
\eeq
Products of tensors  are defined by summation over pairs of indices: 
$({\tens C} {\bf A})_{ij} = C_{ijkl}A_{kl}$ and $({\tens A}{\tens B})_{ijkl} = A_{ijpq}B_{pqkl}$.  Thus, 
${\tens I}{\bd I} = {\tens J}{\bd I} = {\bd I}$, 
${\tens I}{\tens I} = {\tens I}$,  ${\tens J}{\tens J} = {\tens J}$, and 
${\tens I}{\tens J} ={\tens J}{\tens I} = {\tens J}$.

The inner product of a pair of tensors of the same order is   defined 
\beq{002}
\langle u,\, v \rangle = {\rm tr} (uv),  
\eeq
where $\tr {\bf A} = A_{ii}$ and $\tr {\mathbb A} = A_{ijij}$, and the Euclidean norm of a tensor is 
\beq{0021}
\| u\| \equiv 
\langle u,\, u \rangle^{1/2} \, . 
\eeq
This will be used to compare  differences between tensors.  

An anisotropic  elastic stiffness tensor $\tens C$ relates stress $\bf T$ and strain $\bf E$ 
according to the generalized form of Hooke's law:
\beq{00}
{\bf T} = {\tens C}{\bf E}\, . 
\eeq
Stress and strain are symmetric  second order  tensors,   implying that their components are also symmetric,   $T_{ij} = T_{ji}$, $E_{ij} = E_{ji}$,   which in turn implies  the first two of the following identities
  \beq{01}
  C_{ijkl}= C_{ijlk},    \qquad 
  C_{ijkl}= C_{jikl},   \qquad 
  C_{ijkl}= C_{klij} .
  \eeq
 The third property is a consequence of the existence of a positive strain energy of the form 
 $W = \frac12 \langle {\bf E}, \, {\mathbb C}{\bf E}\rangle $, which also  constrains the moduli to be positive definite.  
The squared norm of the elasticity tensor is   
\beq{l1}
\|
{\tens C} \|^2 \equiv \langle {\tens C},\, {\tens C} \rangle 
=  C_{ijkl}C_{ijkl} \, . 
\eeq

%The tensors  ${\tens J}$ and ${\tens K}$ have the multiplication properties ${\tens J}{\tens J} = {\tens J}$, ${\tens K}{\tens K} = {\tens K}$, ${\tens J}{\tens K} = {\tens K}{\tens J} = 0$, and lengths $\|{\tens J}\|^2 = 1$, $\|{\tens K}\|^2 = 5$. 

The 21-dimensional space of elasticity tensors 
 can be decomposed  as a 15-dimensional space of  totally symmetric
fourth order tensors plus a 6-dimensional space of asymmetric  fourth
order tensors \cite{backus,spencer1970}, by
\beq{2.022}
{\tens C} = {{\tens C}}^{(s)} + {{\tens C}}^{(a)}, 
\eeq
where
\beq{03453}
{C}^{(s)}_{ijkl}= \frac13\, (C_{ijkl} + C_{ilkj} + C_{ikjl} ) .  
\eeq
The elements of the totally symmetric part  satisfy the  relations  ${C}^{(s)}_{ijkl}={C}^{(s)}_{ikjl}$ in addition to \rf{01}. Thus, ${C}^{(s)}_{ijkl}$ is unchanged under any rearrangement of its indices, and ${{\tens C}}^{(s)}$ has at most $15$ independent elements.   This can be seen by the explicit representation of the asymmetric part in terms of the remaining six  independent elements, which are the components of the symmetric second order tensor 
\beq{D}
D_{ij} = C_{ijkk} - C_{ikjk}. 
\eeq
Thus, 
\beq{ca}
C^{(a)}_{ijkl}=\frac13\big[ 2D_{ij}I_{kl}+ 2D_{kl}I_{ij} 
- D_{ik}I_{jl}- D_{il}I_{jk} - D_{jk}I_{il} - D_{jl}I_{ik}
+ D_{mm} (I_{ijkl} - 3J_{ijkl})
\big]\, . 
\eeq
An asymmetric tensor satisfies, by definition \cite{backus}, 
\beq{934}
C^{(a)}_{ijkl} + C^{(a)}_{ilkj} + C^{(a)}_{ikjl} =    0.
\eeq
Note that  the symmetric part  
of an asymmetric tensor is zero, 
and {\it vice versa}, and that the decomposition \rf{2.022} is orthogonal in the sense of the Euclidean norm, 
$\| {\tens C} \|^2= \|{\tens C}^{(s)} \|^2 + \|{\tens C}^{(a)} \|^2 $. 
The partition of $\tens C$ as a sum of totally symmetric and traceless asymmetric tensors is the first step in  Backus' harmonic decomposition of elasticity tensors \cite{backus}, a partition that  has proved useful for representations of elastic tensors \cite{baerheim} and has also been used to prove that there are exactly eight distinct elastic symmetries   \cite{FV96}. 
Finally, for future reference, note  that the totally symmetric part of the fourth order identity is 
\beq{0032}
I^{(s)}_{ijkl} =  \frac13\, ( \delta_{ij}\delta_{kl}+ \delta_{ik}\delta_{jl}
+ \delta_{il}\delta_{kj}) 
\qquad \Leftrightarrow \qquad 
{\tens I}^{(s)}  = \frac23 {\tens I} +  {\tens J} \, . 
\eeq 

 \subsection{The acoustical tensor and the tensor ${\tens C}^*$}
 
 The acoustical tensor, also known as Christoffel's matrix \cite{Musgrave},  arises in the study of plane crested waves with displacement of the form ${\bf u}({\bf x},t) = 
  {\bf b}\, h({\bf n}\cdot {\bf x} - vt) $, where $\bf b$ is  a fixed unit vector describing the  polarization,  ${\bf n}$ is the phase direction, also a unit vector,  $v$ is the phase velocity, and $h$ is an arbitrary but sufficiently smooth function.  
Substituting this wave form into the equations of motion
 \beq{588} C_{ijkl}\, \frac{\partial^2 u_k }{\partial x_j\partial x_l} = \rho \ddot u_i , 
 \eeq
 where $\rho $ is the density, implies that it is a solution 
 if and only if ${\bf b}$ and $\rho v^2$ are eigenvector and eigenvalue  of the second order  tensor $\bf Q$, 
\beq{301}
Q_{ik}({\bd n} )  = C_{ijkl}\, n_jn_l\, . 
\eeq
This definition of   the acoustical tensor  is not the product a fourth order tensor with a second order tensor.  In order to express it in this form, which simplifies the analysis later,  
introduce the related fourth order tensor ${\tens C}^*$
  defined by 
\beq{mix1}
 C_{ijkl}^* = \frac12 \big( C_{ikjl} + C_{iljk}\big) \, .    
\eeq
Thus,
\beq{3011}
Q_{ij}({\bd n} )  = C_{ijkl}^*\, n_kn_l \qquad \Leftrightarrow \qquad {\bd Q}({\bd n})  = {\tens C}^*\, {\bd n} \otimes {\bd n} \, .
\eeq

The operation defined by $^*$ is of fundamental importance in solving the generalized Fedorov problem, and therefore some properties   are noted.   First,     $^*$ is a linear operator that commutes with  taking the symmetric and asymmetric parts of a tensor:
${\tens C}^{(s)*}  = {\tens C}^{*(s)}  = {\tens C}^{(s)}$ and $ {\tens C}^{(a)*}   = {\tens C}^{*(a)}   = -\frac12 {\tens C}^{(a)}$.  
Accordingly, ${\tens C}^*$ is partitioned as
\beq{p}
{\tens C}^* = {{\tens C}}^{(s)} -\frac12 {{\tens C}}^{(a)}\, ,  
\eeq
and repeating the $^*$ operation $n$ times  yields ${\tens C}^{n*} = {{\tens C}}^{(s)} +(-1/2)^n \,{{\tens C}}^{(a)}$.  Taking $n=2$ implies the identity 
\beq{006}
{\tens C} =  2 {\tens C}^{**} -  {\tens C}^*,  
\eeq
from  which  ${\tens C}$ can be found from ${\tens C}^*$.  Hence the  mapping $ {\tens C}\leftrightarrow  {\tens C}^*$ is one-to-one  and invertible, that is bijective. This property is important in the inverse problem of determining elastic moduli from acoustic data  \cite{norris89}.   Acoustic wave speeds and associated quantities are  related primarily to ${\tens C}^*$ through the acoustical tensor also known as the Christoffel matrix \cite{Musgrave},  and this  can be determined uniquely  from ${\tens C}$ using \rf{006}.  
Decompositions of ${\tens C} $ and ${\tens C}^*$ into totally symmetric and asymmetric parts are unique, and knowledge of one decomposition  implies the other.  In particular, 
${\tens C} ={\tens C}^*$ if and only if the asymmetric parts of both are zero.  This 
occurs if the moduli satisfy $ C_{ijkk} = C_{ikjk}$, which together with the symmetries \rf{01}  are equivalent to the  Cauchy relations, see Sec. 4.5 of Musgrave \cite{Musgrave}.   We note that the operation $^*$ is self adjoint in the sense that the  following  is true for any pair of elasticity tensors, 
\beq{1071}
\langle {\tens A},\,   {\tens B}^* \rangle = \langle {\tens A}^* ,\,   {\tens B} \rangle\, . 
\eeq

Elastic moduli are usually defined by  the Voigt notation:  $C_{ijkl} \equiv c_{IJ}$, where $I,J = 1,2,3,\ldots 6$ 
and $I=1,2,3,4,5,6$ correspond to $ij=11,22,33,23,13,12$, respectively, i.e. 
%and $c_{IJ}, \ I,J=1,2,\ldots 6$ are the elements of ${\tens C}$ in the Voigt notation, i.e. 
\beq{a1a}
{\tens C}  \quad\leftrightarrow \quad 
{\bd C} =
\begin{pmatrix}
c_{11} & c_{12} & c_{13} & 
 c_{14} &  c_{15} & c_{16} 
\\ & & & & & \\
 & c_{22} & c_{23} & 
 c_{24} &  c_{25} &  c_{26} 
\\ & & & & & \\
 & & c_{33} & 
  c_{34} &  c_{35} &  c_{36} 
\\ & & & & & \\
 &   &   
 & c_{44} & c_{45} & c_{46}
\\ & & & & & \\
   S &Y  &M & & c_{55} & c_{56}
\\ & & & & & \\
&&&& & c_{66}
\end{pmatrix} . 
\eeq
The components of ${\tens C}^*$
are then 
\beq{2.03}
 {\tens C}^*  \quad\leftrightarrow \quad 
 {\bd C}^* =
\begin{pmatrix}
~~c_{11}~~ & ~~c_{66}~~ & ~~c_{55}~~ & c_{56} & c_{15} & c_{16} \\
~ & ~ & ~ & ~ & ~ & ~  \\ 
 ~    & c_{22} & c_{44} & c_{24} & c_{46} & c_{26} \\ 
~ & ~ & ~ & ~ & ~ & ~  \\ 
   ~  &   ~     & c_{33} & c_{34} & c_{35} & c_{45} \\ 
~ & ~ & ~ & ~ & ~ & ~  \\ 
  ~  &  ~ & ~ & \frac12 (c_{44}+c_{23}) 
                & \frac12 (c_{45}+c_{36} )
                & \frac12 (c_{46}+c_{25})  \\
~ & ~ & ~ & ~ & ~ & ~  \\ 
  S  &  Y & M & ~& \frac12 (c_{55}+c_{13} )
                 & \frac12 (c_{56}+c_{14} ) \\
~ & ~ & ~ & ~ & ~ & ~  \\ 
~ & ~ & ~ & ~ & ~ & \frac12 (c_{66}+c_{12} )
\end{pmatrix} . 
\eeq

\section{Fedorov's problem for particular symmetries}\label{sec3}

\subsection{Definition of the problem}
We assume as known  the elasticity tensor $\tens C$ of arbitrary material symmetry with  as many as  21 independent components.  A particular material symmetry is chosen with prescribed symmetry axes or planes. For instance,  transverse isotropy with symmetry axis in the direction ${\bf a}$, or cubic symmetry with orthogonal cube axes  ${\bf a}, {\bf b}, {\bf c}$. 
Fedorov's problem for particular symmetries is to determine the elastic stiffness  ${\tens C}_{Sym}$ of the chosen material symmetry which is the best fit 
in the sense that it minimizes the orientation averaged squared difference of the acoustical tensors. We introduce the  acoustical distance function 
\beq{fed1}
f ({\tens C}, {\tens C}_{Sym} ) \equiv \frac1{4\pi}\, \int\limits_0^{2\pi} \mbox{d} \phi \int\limits_0^{\pi} \sin\theta \mbox{d} \theta   \, \|{\bd Q}({\tens C} , {\bd n}) - {\bd Q}({\tens C}_{Sym} , {\bd n}) \|^2 \, ,  
\eeq
with ${\bd n} = \sin\theta (\cos\phi {\bd e}_1 +\sin\phi {\bd e}_2)+\cos\theta {\bd e}_3$. 
 The same function was considered by Fedorov \cite{fed} with ${\tens C}_{Sym}$ restricted to isotropic elasticity.  Thus, substituting  ${\tens C}_{Sym} = \alpha_1 {\tens I} + \alpha_2 {\tens J}$ into \rf{fed1} one obtains, after simplification, a positive definite quadratic in the two unknowns $\alpha_1 $ and $\alpha_2$.  A simple minimization yields $\alpha_1 = 2\mu$ and $\alpha_2 = 3\kappa -2\mu$ where $\kappa$ and $\mu$ are defined in eq. \rf{ie}.  
The  symmetry of  ${\tens C}_{Sym}$ can be considered as arbitrary for  the general problem addressed here,  although Fedorov's isotropic result is recovered as a special case of the  general solution discussed in   Section \ref{ex}. 

The integral over directions in \rf{fed1} can be removed using 
\beq{fed3}
\frac1{4\pi} \int\limits_0^{2\pi} \mbox{d} \phi \int\limits_0^{\pi} \sin\theta \mbox{d} \theta   \, {\bd n}\otimes{\bd n}\otimes{\bd n}\otimes{\bd n}  = \frac15\, {\tens I}^{(s)}, 
\eeq
where ${\tens I}^{(s)}$ is defined in  eq. \rf{0032}. 
The identity follows by noting that the  integral must be a totally symmetric isotropic fourth order tensor of the form $a  {\tens I}^{(s)}$.  Taking the trace of both sides and  using 
$\tr {\tens I}^{(s)}  = I^{(s)}_{ijij} = 5$, $\tr ({\bd n}\otimes{\bd n}\otimes{\bd n}\otimes{\bd n}) = 1$,  gives $a=1/5$. 
Thus, since 
\beq{fed2}
{\bd Q}({\tens C} , {\bd n})  =  {\tens C}^*  \, {\bd n}\otimes{\bd n} ,\qquad
{\bd Q}({\tens C}_{Sym} , {\bd n})  = {\tens C}_{Sym}^* \, {\bd n}\otimes{\bd n} ,  
\eeq
 the distance function   reduces  to 
\beq{fed4}
f = \frac15\, 
\langle ({\tens C}^* - {\tens C}_{Sym}^*),\,  {\tens I}^{(s)} ({\tens C}^* - {\tens C}_{Sym}^*)\rangle \, . 
\eeq
Define the modified inner product for elasticity tensors: 
\beq{007}
\langle {\tens A},\,   {\tens B} \rangle_a
\equiv {\rm tr} ( {\tens I}^{(s)} {\tens A}  {\tens B} ) 
= 
\langle {\tens A},\,  {\tens I}^{(s)} {\tens B} \rangle \, , 
\eeq
and norm 
\beq{0071}
\| {\tens A}\|_a = \langle {\tens A},\,   {\tens A} \rangle_a^{1/2}
\, . 
\eeq
Then Fedorov's problem for particular symmetries amounts to: 
\beq{008}
 \mbox{ Fedorov  } \quad \Leftrightarrow 
 \quad \mbox{ minimize  }  \|{\tens C}^* - {\tens C}_{Sym}^*\|_a \, . 
 \eeq

The reason  the problem is expressed in this form   is to make the connection with the notion of projection onto the chosen elastic symmetry, or equivalently of finding the elastic tensor of the chosen symmetry nearest to the given elasticity $\tens C$.  As mentioned in the Introduction, this question has been addressed by several authors and has an explicit solution.  The issue is to find the elastic tensor ${\tens C}_{Sym}$ which minimizes the Euclidean distance function $d$, where
\beq{132}
d({\tens C}, {\tens C}_{Sym}) \equiv  \|{\tens C} - {\tens C}_{Sym}\|\, . 
\eeq
Comparing eqs. \rf{008} and \rf{132}, the two problems appear tantalizingly similar.  It should be realized that Fedorov's problem involves ${\tens C}^*$ not ${\tens C}$, and that the norms are distinct.  However, it will be proved  that the problems share the same solution: 
 \beq{0082}
 \{ {\tens C}_{Sym}:\,  \min\limits_{{\tens C}_{Sym} } \,  \|{\tens C}^* - {\tens C}_{Sym}^*\|_a \}
  \equiv 
 \{ {\tens C}_{Sym}:\,  \min\limits_{{\tens C}_{Sym} } \, \|{\tens C} - {\tens C}_{Sym}\| \}\, , 
 \eeq
which gives the  central result of this paper, i.e. 
 \beq{0081}
 \mbox{ Fedorov  } \quad \Leftrightarrow \quad 
 \mbox{ Euclidean projection.} 
 \eeq
This equivalence enables us to provide an explicit solution to the generalized Fedorov problem, e.g. using the methods of Gazis et al. \cite{Gazis63}, Browaeys and Chevrot \cite{Browaeys04} or others. 
The remainder of this Section develops a proof by construction of the solution of the Fedorov problem, which is shown to be identical to the Euclidean projection.   Some further concepts and notation are required and introduced next.

 \subsection{Basis tensors}
\footnote{}
The solution uses a fundamental decomposition  of the chosen material symmetry using basis tensors.  These form a vector space for the symmetry in the sense that any elasticity tensor of that symmetry may be expressed uniquely in terms of $N$ linearly independent tensors ${\tens V}_1, {\tens V}_2, 
\ldots {\tens V}_N$, where $2\le N\le 13 $ is the dimension of the vector space for the material symmetry.  
For instance,  isotropic elasticity tensors are of the form ${\tens C}_{Iso} = \alpha_1 {\tens I} + \alpha_2 {\tens J}$, $\alpha_1, \alpha_2 >0$.  The  procedure is  analogous for other material symmetries, cubic, transversely isotropic, etc., and is described in detail by Walpole \cite{walpole84} who provides expressions for the base tensors of the various symmetries.  Thus,  $N=2,\, 3, \, 5,\, 9$ for isotropy, cubic symmetry,  transverse isotropy and orthorhombic symmetry, respectively.  $N=13$ corresponds to monoclinic, which is the lowest symmetry apart from triclinic   (technically $N=21$) which is no symmetry. 
The precise form of the basis tensors is irrelevant here (examples are given in Section \ref{ex},  full details are in \cite{walpole84},  and Kunin \cite{Kunin} develops a similar tensorial decomposition), all that is required is that  they be linearly independent, and consequently any tensor with the desired symmetry  can be written    
\beq{545}
{\tens C}_{Sym}  = \sum\limits_{i=1}^N \, \beta_i {\tens V}_i\, .  
\eeq
The coefficients follow by taking inner products with the basis tensors.   Let  $\bd{\Lambda}$ be the  the $N\times N$  symmetric matrix  with elements   
 \beq{fed7}
  \Lambda_{ij}   \equiv  \langle{\tens V}_i , \,  {\tens V}_j \rangle \, .   
 \eeq
$\bd{\Lambda} $ is  invertible by virtue of the linear independence of the basis tensors, and therefore
\beq{546}
\beta_i = \sum\limits_{j=1}^N \, \Lambda_{ij}^{-1} \langle {\tens C}_{Sym},\, {\tens V}_j \rangle
    \, .  
\eeq
Conversely, eq. \rf{545} describes all possible tensors with the given symmetry, and in particular, 
${\tens C}_{Sym}  = 0$ if and only if $\beta_i=0$, $i=1,2,\ldots N$.

 \subsection{The elastic projection}
 
 It helps to first derive the solution that minimizes the Euclidean distance function  of \rf{132}.  
By expressing the   unknown projected solution in the form \rf{545}, it follows that 
the minimum of $\|{\tens C} - {\tens C}_{Sym}\|^2$ is determined by setting to zero the derivatives with respect to $\beta_i$, which gives the system of simultaneous equations
\beq{fed61}
 \sum\limits_{j=1}^N \langle{\tens V}_i , \,  {\tens V}_j \rangle \,\beta_j    = \langle{\tens V}_i , \,  {\tens C} \rangle ,  \qquad i = 1,2,\ldots N.   
 \eeq
The inner products are the elements of the invertible matrix $\bd{\Lambda} $, and so  
\beq{504}
{\tens C}_{Sym}  = \sum\limits_{i, j =1}^N  \langle{\tens C} , \,  {\tens V}_i \rangle  \, 
\,  \Lambda_{ij}^{-1} \, {\tens V}_j  \, . 
\eeq 
Furthermore,  the distance function at the optimal ${\tens C}_{Sym}$ satisfies 
\beq{843}
\|{\tens C} - {\tens C}_{Sym}\|^2 = \|{\tens C} \|^2 - \| {\tens C}_{Sym}\|^2 \, , 
\eeq
as expected for a projection using the Euclidean norm. This is essentially the method used by Arts et al. \cite{Arts91}. Helbig \cite{Helbig95} describes the procedure as the 'optimum approximation'   of an arbitrary elasticity tensor  
  by projection of a 21-dimensional  vector onto a subspace of fewer dimensions, and Browaeys and Chevrot  \cite{Browaeys04} list the explicit forms of the 21D projection operators for various symmetries.  

Define the projection operator ${\cal P}_{Sym}$ which maps ${\tens C}$ onto the chosen symmetry,
\beq{505}
{\cal P}_{Sym} {\tens C}  \equiv  \sum\limits_{i, j =1}^N  \langle{\tens C} , \,  {\tens V}_i \rangle  \, 
\,  \Lambda_{ij}^{-1} \, {\tens V}_j  \, . 
\eeq 
Equations \rf{504} through \rf{505} imply that ${\cal P}_{Sym} {\tens C} $ is the Euclidean projection, also equal to the closest elasticity tensor of the chosen symmetry to ${\tens C}$.   We note the following important property:
\beq{506}
{\cal P}_{Sym} {\tens C}^*  = \big( {\cal P}_{Sym} {\tens C} \big)^*\, . 
\eeq
In other words, the operation $^*$ commutes with the projection operator.  This is not surprising if one considers that the $^*$ operation is a linear mapping on the symmetric and asymmetric parts of of $\tens C$,  and therefore $^*$ the maintains the material symmetry of ${\tens C}$.  However, a more  detailed   proof of the identity \rf{506} is provided in the Appendix.

 \subsection{Solution of the generalized Fedorov problem}
 
We now  calculate the optimal ${\tens C}_{Sym}$ for the generalized Fedorov problem, and show that it is equivalent to the moduli from  the Euclidean norm.  The starting point this time is to express the unknown ${\tens C}_{Sym}^*$  (rather than ${\tens C}_{Sym}$) in terms of the basis tensors,
\beq{fed5}
{\tens C}_{Sym}^*  = \sum\limits_{i=1}^N \, \alpha_i {\tens V}_i
%\qquad \Leftrightarrow \qquad
\, . 
\eeq 
This is justified by the fact that all tensors of the given symmetry are linear combinations of the basis tensors.  Furthermore, the coefficients  $\alpha_i$ are related to those in eq.  \rf{545} by the $N\times N$ matrix 
$\bd P$ introduced in the Appendix.  Let
$\bd{\alpha}$  and $\bd{\beta}$  denote the $N\times 1$ arrays with elements $\alpha_i$ and $\alpha_i$, then 
\beq{324}
 \bd{\beta}  = \bd{P}^t \bd{\alpha}
 \qquad \Leftrightarrow \qquad
 \bd{\alpha} = 2\bd{P}^t \bd{\beta} - \bd{\beta}. 
 \eeq
Equation \rf{008} implies that the coefficients coefficients $\alpha_1,\, \alpha_2, \ldots \alpha_N$  satisfy the system of $N$ equations 
\beq{fed6}
 \sum\limits_{j=1}^N \langle{\tens V}_i , \,  {\tens V}_j \rangle_a \,\alpha_j    = \langle {\tens V}_i , \,  {\tens C}^* \rangle_a ,  \qquad i = 1,2,\ldots N, 
 \eeq
or, in matrix format, 
 \beq{3241}
 {\bf S} \bd{\Lambda} \bd{\alpha}   =   {\bf S}\bd{\gamma}\, .
 \eeq
Here    
   $\bd{\gamma}$ with elements $\gamma_i$ and   the $N\times N$   matrix $\bd S $ are defined by 
   \begin{subequations}
 \beqa{323}
 \gamma_i    &=& \langle{\tens V}_i , \,  {\tens C}^* \rangle ,  
\\
 \sum\limits_{j=1}^N S_{ij} {\tens V}_j &=&   \frac12 ({\tens I}^{(s)}{\tens V}_i + {\tens V}_i {\tens I}^{(s)}) \equiv {\tens U}_i
 \, .  
 \eeqa
 \end{subequations}
 
 The tensors ${\tens U}_i$, $i=1,2,\ldots N$  form a linearly independent set of basis tensors for the given symmetry.  This may be seen by assuming the contrary, i.e. that there is a set of non-zero coefficients $a_i$ such that 
 \beq{2345}
 \sum\limits_{i=1}^N a_i {\tens U}_j  = 0. 
 \eeq
 Let ${\tens A}$ be the non-zero tensor
 \beq{231}{\tens A} = 
 \sum\limits_{i=1}^N a_i {\tens V}_j  ,
 \eeq
 then eq. \rf{2345} requires  that 
  \beq{232}
  \frac23 {\tens A}  
    + \frac12 ({\tens A} {\bd I} ) \otimes {\bd I} 
    + \frac12 {\bd I} \otimes({\tens A} {\bd I} ) = 0. 
 \eeq
 Multiplication by ${\bd I}$ implies
 \beq{233}
  \frac{13}{6} {\tens A} {\bd I} 
  + \frac32  \langle {\tens J},\, {\tens A}\rangle  {\bd I}
   = 0, 
 \eeq
 and taking the inner product with ${\bd I}$ gives 
 \beq{234}
 11 \, \langle {\tens J},\, {\tens A}\rangle  
   = 0.  
 \eeq
 Therefore, 
 \beq{235}
 {\tens A} = 0,
 \eeq
 and  the ${\tens U}_i$  tensors form a linearly independent basis for the symmetry. In particular, the tensors ${\tens V}_i$ can be expressed in terms of this alternate basis, and so   the matrix 
 ${\bf S} $  is  invertible.  Hence  
  \beq{325}
  \bd{\alpha}   =   \bd{\Lambda}^{-1}\bd{\gamma}\, . 
 \eeq
 
We are now in a position to determine the optimal ${\tens C}_{Sym}^*$ and hence ${\tens C}_{Sym}$.  
Equations \rf{fed5} and \rf{325}, along with   \rf{505},   imply 
\beqa{fed52}
{\tens C}_{Sym}^*  &=& \sum\limits_{i, j=1}^N    \langle {\tens C}^* ,\, {\tens V}_i   \rangle  \Lambda_{ij}^{-1}  {\tens V}_j
\nonumber \\
&= & {\cal P}_{Sym} {\tens C}^* \, . 
\eeqa
Using the fundamental property of the projection operator \rf{506}, the optimal elasticity as determined by \rf{fed52} is seen to be exactly the same as the Euclidean projection, i.e. of eq. \rf{504}.   
This completes the proof of the main result, the equivalence \rf{0081}. 

\section{Examples}\label{ex}

The general procedure for projection is illustrated for several symmetries, and an example application is discussed in this Section.  

\subsection{Basis functions for isotropic, cubic and hexagonal materials}

The tensor decomposition procedure is described  for the three highest symmetries: isotropic, cubic and transversely isotropic.   The fundamental matrices $\bd \Lambda $ and $\bd P$ are given explicitly.  

\subsubsection{Isotropic approximation $(N=2)$}

Let the basis tensors be  ${\tens V}_1 = {\tens J}$, ${\tens V}_2 = {\tens K} \equiv {\tens I} -{\tens J}$.  Then $\bd{\Lambda} $ is a $2\times 2$ diagonal matrix, $\bd{\Lambda} = \diag  ( 1,\, 5)$ and 
the optimal moduli are  
\beq{106}
{\tens C}_{Iso}  = 3\kappa\,{\tens J} + 2\mu \, {\tens K} , 
\eeq
where 
\beq{110}
\kappa  =\frac13 \langle {\tens J}  ,\, {\tens C} \rangle,
\qquad
\mu = \frac{1}{10}\langle {\tens K} ,\, {\tens C} \rangle \, . 
\eeq 
This is precisely Fedorov's original result \cite{fed}, eq. \rf{ie}. The $^*$ operation is defined by the matrix $\bf P$ of eq. $\rf{320}_1$, which is 
\beq{is1}
{\bf P } = \begin{pmatrix}
~\frac13~ & ~\frac13~ \\ & \\
\frac53 & \frac16 
\end{pmatrix} \, . 
\eeq

\subsubsection{Cubic materials $(N=3)$}
Let ${\bf a}, {\bf b},{\bf c}$ be the cube axes, and select as basis tensors \cite{walpole84} 
${\tens V}_1 = {\tens J}$, ${\tens V}_2 = {\tens L} \equiv {\tens I} - {\tens H}$, and ${\tens V}_3 = {\tens M} \equiv {\tens H}-{\tens J} $, 
where 
\beq{53}
{\tens H} = {\bf a} \otimes{\bf a}\otimes{\bf a}\otimes{\bf a} +{\bf b}\otimes{\bf b}\otimes{\bf b}\otimes{\bf b}+ {\bf c}\otimes{\bf c}\otimes{\bf c}\otimes{\bf c}\, . 
\eeq
%Note that ${\tens J}$, $ {\tens L}$ and ${\tens M}$ are orthogonal and idempotent under multiplication, and    ${\tens I} = {\tens J} + {\tens L} + {\tens M}$.  
Then $\bd{\Lambda} = \diag ( 1,\, 3,\, 2)$ 
and the optimal moduli of cubic symmetry are 
\beq{116}
{\tens C}_{Cub}  = 3\kappa '\,{\tens J} + 2\mu ' \, {\tens L} + 2\eta \, {\tens M} , 
\eeq
where 
\beq{120}
\kappa ' =\frac13 \langle {\tens J}  ,\, {\tens C} \rangle,
\qquad
\mu '= \frac16\langle {\tens L} ,\, {\tens C} \rangle 
\qquad
\eta = \frac14\langle {\tens M} ,\, {\tens C} \rangle \, . 
\eeq 
It is assumed here that the axes of the cubic material are known, otherwise a numerical search must be performed to find the axes which give the closest, i.e. largest, projection.   This additional step is discussed in the numerical example below. 
Also, 
\beq{c1}
{\bf P } = \begin{pmatrix}
~ \frac13 ~&~ \frac13 ~&~ \frac13 ~ \\ && \\ 
1& \frac12 & -\frac12  \\  && \\  
 \frac23 & -\frac13  & \frac23 
\end{pmatrix} \, . 
\eeq

 \subsubsection{Transverse isotropy/hexagonal symmetry $(N=5)$}
  
Let ${\bf a}$ be  the  direction of the symmetry axis, and define \cite{walpole84}
\beq{t4}
{\tens V}_{1}  ={\bd A}  \otimes{\bd A} , \quad
{\tens V}_{2}  =\tfrac12 {\bd B} \otimes{\bd B} ,  \quad
{\tens V}_{3}=  \tfrac12 ( {\bd A}  \otimes{\bd B} + {\bd B}  \otimes{\bd A}) ,\quad
{\tens V}_{4} =  2 {\tens V}_{3}^* , \quad 
{\tens V}_{5} =  2 {\tens V}_{2}^* -{\tens V}_{2}, \quad
\eeq
where ${\bd A} = {\bf a}\otimes {\bf a}$ and ${\bd B} = {\bd I} - {\bd A} $. 
Then $\bd{\Lambda} = \diag ( 1,\, 1,\, 1,\, 2,\, 2)$ 
 %Also, we note that \cite{walpole84,c2}\beq{t9} {\tens I} = {\tens V}_1 + {\tens V}_2 + {\tens V}_4+{\tens V}_5\, , \qquad {\tens J} = ( {\tens V}_1 + 2{\tens V}_2 + 2{\tens V}_3)/3 \, . \eeq
and the  optimal TI moduli are given by 
\beq{t10}
 {\tens C}_{TI} = \sum\limits_{i=1}^3 \langle {\tens C},\, {\tens V}_i\rangle\, {\tens V}_i
 + \frac12 \sum\limits_{i=4}^5 \langle {\tens C},\, {\tens V}_i\rangle\, {\tens V}_i\, .  
\eeq
Also, 
\beq{877}
{\bf P } = \begin{pmatrix}
~ 1 ~&~ 0~&~ 0 ~ &~ 0 ~ &~ 0  ~\\ &&&&  \\ 
0& \frac12 & 0 & 0 &  \frac12  \\  &&&& \\  
0& 0 & 0 & \frac12 &  0  \\  &&&& \\  
0& 0 & 1 & \frac12 &  0  \\  &&&& \\   
0&  1 & 0 & 0 & 0
\end{pmatrix}\, .
\eeq

\subsection{Application to acoustically measured data}

Let us assume that the material has cubic symmetry  but the cube axes orientations    
%${\bd a}$, ${\bd b}$, ${\bd c}$, 
are unknown.   The effective cubic moduli are, in coordinates coincident with the cube axes,   
\beq{639}
{\bd C}_{Cub}  = \begin{pmatrix} 
c_{11}^c & c_{12}^c & c_{12}^c &0 & 0 & 0  \\ & & & & & \\
c_{12}^c & c_{11}^c & c_{12}^c & 0 & 0 & 0  \\ & & & & & \\
c_{12}^c & c_{12}^c & c_{11}^c & 0 & 0 & 0   \\ & & & & & \\
0 & 0 & 0 & c_{66}^c & 0&  0   \\ & & & & & \\
0 & 0 & 0 & 0 & c_{66}^c & 0    \\ & & & & & \\
0 & 0 & 0 & 0 & 0 & c_{66}^c 
\end{pmatrix} , 
\eeq
where the three constants can be expressed in terms of the bulk modulus and the two shear moduli,
\beq{1204}
c_{11}^c = \kappa ' + \frac43 \eta, 
\qquad 
c_{12}^c = \kappa ' - \frac23 \eta, 
\qquad 
c_{66}^c = \mu '\, .  
\eeq
The effective bulk modulus of \rf{120} is an isotropic invariant that is independent of the cube axes orientation, 
\beq{1201}
\kappa ' =\frac19 C_{iijj} = \frac19 \big(
c_{11}+c_{22}+c_{33}+2c_{12}+2c_{23}+2c_{31}\big) \, . %= 170.11 \quad {\rm GPa}. 
\eeq
Similarly, the combination $(3\mu ' + 2\eta)/5 = \mu$, the isotropic shear modulus, implying 
\beq{1202}
6\mu ' + 4\eta = C_{ijij} - \frac13 C_{iijj} = 
\frac23 \big( c_{11}+c_{22}+c_{33}-c_{12}-c_{23}-c_{31}\big) 
+2\big( c_{44}+c_{55}+c_{66}\big) \, . %=  968.7\quad {\rm GPa}. 
\eeq
Only one of the three cubic parameters depends upon the orientation of the axes, and it  follows by considering the squared length of the projected elastic tensor,  
\beq{1203}
\| {\tens C}_{Cub}\|^2 = 9 {\kappa '}^2 + 12 {\mu '}^2 + 8\eta^2
= 9 {\kappa '}^2 + \frac15 (6\mu ' + 4\eta)^2+ \frac65 (2\mu ' -2\eta)^2 \, .
\eeq
The first two terms are independent of the cube axes orientation, and therefore the closest cubic projection  maximizes the final term.  Note that 
\beq{1205}
2\mu ' -2\eta = \frac13C_{ijij} + \frac16 C_{iijj} -\frac56 \langle {\tens C}, {\tens H}\rangle = 
\frac52 \kappa ' + 2\mu ' +\frac43 \eta -\frac56 \langle {\tens C}, {\tens H}\rangle \, . 
\eeq
The moduli $\kappa '$, $\mu '$ and $\eta$ must all be positive in order for the material to have positive definite strain energy.  Therefore, $(2\mu ' -2\eta)^2$ is  maximum when 
$\langle {\tens C}, {\tens H}\rangle $ is minimum.  The latter is also a positive quantity, which may be expressed
\beq{54}
\langle {\tens C}, {\tens H}\rangle = 
C_{ijkl}H_{ijkl} = c_{11}' + c_{22}' + c_{33}'   \, , 
\eeq
where $c_{IJ}'$  are the elements of ${\tens C} $ of \rf{638} in the coordinate system coincident with the cube axes. 

We can also write 
\beq{541}
\langle {\tens C}, {\tens H}\rangle =  
\tilde{\bd a}^t{\bd C}\tilde{\bd a} +
\tilde{\bd b}^t{\bd C}\tilde{\bd b} +
\tilde{\bd c}^t{\bd C}\tilde{\bd c} \, , 
\eeq
where 
\beq{542}
 \tilde{\bd a}^t  = \big(a_1^2 ,\,   a_2^2 ,\,    a_3^2 ,\,  
2 a_2a_3 ,\,  2 a_3a_1 ,\,    2 a_1a_2 \big),\quad etc.  
\eeq
The minimum can be found by numerically 
searching over all possible orientations using Euler angles  to transform from the fixed vector basis to the cube axes.   The rotated (cube) axes are simply the columns of the $3\times 3$ transformation matrix ${\bd R}\in SO(3)$. 

As a specific application we reconsider the elastic moduli obtained from ultrasonic measurements by Francois et al. \cite{fgm}.  In the notation of eq. \rf{a1a}, the raw data for the stiffness tensor  is\footnote{The element $c_{53}$ is given as $49$ in \cite{fgm}, which appears to be a typographical error.} 
\beq{638}
{\bd C}  = \begin{pmatrix} 
243 & 136 & 135 & 22 & 52 & -17  \\ & & & & & \\
136 & 239 & 137 & -28 & 11 & 16  \\ & & & & & \\
135 & 137 & 233 & 29 & -49 & 3   \\ & & & & & \\
22 & -28 & 29 & 133 & -10&  -4   \\ & & & & & \\
52 & 11 & -49 & -10 & 119 & -2    \\ & & & & & \\
-17 & 16 & 3 & -4 & -2 & 130 
\end{pmatrix} \quad ({\rm GPa}). 
\eeq
We find that  
\beq{35}
{\bd C}_{Cub}  = \begin{pmatrix} 
213.4 & 148.5 & 148.5 & 0 & 0 & 0  \\ & & & & & \\
148.5 & 213.4 & 148.5 & 0 & 0 & 0  \\ & & & & & \\
148.5 & 148.5 & 213.4 & 0 & 0 & 0   \\ & & & & & \\
0 & 0 & 0 & 139.8 & 0&  0   \\ & & & & & \\
0 & 0 & 0 & 0 & 139.8 & 0    \\ & & & & & \\
0 & 0 & 0 & 0 & 0 & 139.8 
\end{pmatrix} \quad ({\rm GPa}),
\eeq
in agreement with Francois et al. \cite{fgm}.  The rotation from the coordinate system $\{ {\bd e}_1,{\bd e}_2,{\bd e}_3\}$ 
of \rf{638} can be expressed as a  rotation  about a single axis $\bd n$ through angle $\theta$ using Euler's theorem  \cite{Baruh}.  The angle and axis  follow from  $    2 \cos \theta =\tr {\bd R}  -1$  and $2\sin \theta \, {\bd n} = \skew {\bd R}$, where $\skew {\bd Y} = -\epsilon_{ijk}Y_{ij}\, {\bd e}_k  $ and $\epsilon_{ijk}$ is the third order alternating tensor.  Thus, we find 
$\theta = 96^\circ$ and ${\bd n} = \big( 0.18,\,     0.06,\,  0.98 \big) $, 
and the full set of moduli in the rotated frame are  
\beq{315}
{\bd C}'  = \begin{pmatrix} 
228  & 141 &  148  & -5  & -1 &  -2  \\ & & & & & \\
141  &  209 &  156  &   -6 &   23 & -1 \\ & & & & & \\
148 &  156 &  203  &   -7  &  -2 &  4  \\ & & & & & \\
-5  &   -6  &   -7 &  144 &    12 &  -3  \\ & & & & & \\
-1  &  23  &  -2  & 12  &  139 &  11 \\ & & & & & \\
-2  &   -1 &   4  &   -3   &   11  & 136
\end{pmatrix}  \quad ({\rm GPa}). 
\eeq
This set of cube axes is unique within rotation under the group of transformations congruent with cubic symmetry.  In this preferred coordinate system, the projection onto the cubic moduli in this frame is simply
\beqa{104}
c_{11}^c &=& \tfrac13 \big( c_{11}'+c_{22}'+c_{33}'\big) = 213.4 \quad ({\rm GPa}),
\nonumber \\ 
c_{12}^c &=& \tfrac13 \big( c_{12}'+c_{23}'+c_{31}'\big)= 148.5,
 \\ 
c_{66}^c &=& \tfrac13 \big( c_{44}'+c_{55}'+c_{66}'\big)= 139.8 . 
\nonumber 
\eeqa
    
\section{Conclusion}

The main result of   this paper is the proof that the Euclidean projection  of anisotropic elastic constants onto a higher material is identical to  minimizing the mean square difference of the slowness surfaces.   This provides a well grounded  acoustical basis for using the Euclidean projection as a natural way to simplify ultrasonic or acoustic data. The equivalence generalizes the 
original result of Fedorov for the best  isotropic acoustical fit to a given set of 
anisotropic moduli. 

\section*{Acknowledgment}   Discussions with Dr. M. Moakher have  benefited this paper and are appreciated. 
    
%^^^^^^^^^^^^^^^^^^^^^^^^^^^^^^^^^^^^^^^^^^^^^^^^^^^^^^^^^^^^^^^^^^
\noindent
\appendix  

\Appendix{The matrix $\bd P$ and proof of eq. \rf{506}  }\label{ap1}

The tensor  ${\tens V}_i^* $ can be expressed using eq. \rf{p} in terms of the totally symmetric and asymmetric parts of ${\tens V}_i$. Both ${\tens V}_i^{(s)}$ and ${\tens V}_i^{(a)}$ possess the same material symmetry as ${\tens V}_i$, and hence the symmetry is inherited by ${\tens V}_i^*$.  Since the ${\tens V}_i$ themselves form a basis for the material symmetry, it follows that ${\tens V}_i^* $ can be written as a linear combination of them.  
Let $\bf P$  be the $N\times N$ matrix  which  defines the $^*$ operation in terms of the basis tensors, that is,   
\beq{320}
{\tens V}_i^* = \sum\limits_{j=1}^N P_{ij} {\tens V}_j . 
\eeq
It follows from eq. \rf{006} that
\beq{321}
{\tens V}_i = - {\tens V}_i^* + 2\sum\limits_{j=1}^N P_{ij} {\tens V}_j^* ,  
\eeq
and hence the inverse of $\bf P$ is 
\beq{pe}
{\bf P}^{-1} = 2 {\bf P}  - {\bf I}_{(N)}, 
\eeq
where ${\bf I}_{(N)}$ is the $N\times N$ identity. 

Consider the identity
\beq{m1}
\langle   {\tens V}_i^* ,\, {\tens V}_j \rangle 
= \langle   {\tens V}_i ,\, {\tens V}_j ^* \rangle, 
\eeq
which follows from  \rf{1071}.  Using $\rf{320}_1$ to eliminate $ {\tens V}_i^*$ and $ {\tens V}_j^*$ from the left and right members, eq. \rf{m1} implies 
\beq{mm2}
{\bf P}\bd{\Lambda} =  \bd{\Lambda}{\bf P}^t\, .  
\eeq
It is worth noting the very specific nature of the matrix $\bd P$ that is required to satisfy eq. \rf{mm2}.  This matrix essentially defines the $^*$ operator, from which the totally symmetric and asymmetric parts of a tensor can be found in terms of the basis tensors. Some examples of $\bf P$ are given in  Section \ref{ex}.

We now turn to the proof of eq. \rf{506}.  
Starting with the definition of ${\cal P}_{Sym}$ in \rf{505}, we have
\beqa{fed53}
 {\cal P}_{Sym} {\tens C}^*
 &=& \sum\limits_{i, j=1}^N    \langle  {\tens C}^* ,\, {\tens V}_i \rangle \Lambda_{ij}^{-1}  {\tens V}_j
 \nonumber \\ 
 &=& \sum\limits_{i, j=1}^N    \langle  {\tens C},\, {\tens V}_i^* \rangle \Lambda_{ij}^{-1}  {\tens V}_j
\nonumber \\ 
 &=& \sum\limits_{i, j =1}^N  \langle{\tens C} , \,  {\tens V}_i \rangle  \, 
\,  X_{ij} \, {\tens V}_j^*
 \, ,
\eeqa
where  \rf{1071} has been used in the second line, and the matrix ${\bf X}$ is defined as 
\beq{644}
{\bf X} = {\bf P}^t \bd{\Lambda}^{-1} {\bf P}^{-1}. 
\eeq
Comparison with $\rf{320}_2$ gives
\beq{355}
 {\bf X} =  \bd{\Lambda}^{-1} \, , 
 \eeq
and substituting from \rf{355} into \rf{fed53} and  comparing it with \rf{504} implies the fundamental property \rf{506}.

%\fontsize{11}{12} \selectfont 
\newpage
%\bibliographystyle{unsrt}%plain}acm}%
%\bibliography{../bib/thermoelastic}
%\end{document}
%%%%%%%%%%%%%%%%%%%%%%%%%%%%%%%%%%%%%%%%%%%%EXTRA

\end{document}
%%%%%%%%%%%%%%%%%%%%%%%%%%%%%%%%%%%%%%%%%%%%EXTRA

for pdf 

%%%%%%%%%%%%%%%%%%%%%%%%%%%%%%%%%%%%%%%%%
\usepackage{color}
\newif\ifpdf
\ifx\pdfoutput\undefined
      \pdffalse
\else
      \pdftrue
\fi
 
\ifpdf
    \pdfcatalog { /PageMode (/UseNone)
                  /OpenAction (fitbh)
    }
 
  \usepackage[pdftex]{graphicx}
  \pdfcompresslevel=9
 
  \usepackage[pdftex]{hyperref}
  \hypersetup{
   pdftitle={Fedorov's problem},
    pdfsubject={acoustics},
    pdfkeywords={elasticity,waves,seismics,ultrasonics},
    pdfauthor={Andrew N. Norris,
               <norris@rutgers.edu>},
    pdfpagemode={UseOutlines},
    bookmarksopen=true,
 %   pagebackref=true,
    colorlinks=true,
    urlcolor=rltblue,
    filecolor=rltgreen,
    linkcolor=rltred,
    citecolor=blue,
    pagecolor=red,
    urlcolor=cyan
  }
  \definecolor{rltred}{rgb}{0.75,0,0}
  \definecolor{rltgreen}{rgb}{0,0.5,0}
  \definecolor{rltblue}{rgb}{0,0,0.75}
\else
  \usepackage{graphicx}
\fi

\if@mathematic
   \def\bd#1{\ensuremath{\mathchoice
                     {\mbox{\boldmath$\displaystyle\mathbf{#1}$}}
                     {\mbox{\boldmath$\textstyle\mathbf{#1}$}}
                     {\mbox{\boldmath$\scriptstyle\mathbf{#1}$}}
                     {\mbox{\boldmath$\scriptscriptstyle\mathbf{#1}$}}}}
\else
   \def\bd#1{\ensuremath{\mathchoice
                     {\mbox{\boldmath$\displaystyle#1$}}
                     {\mbox{\boldmath$\textstyle#1$}}
                     {\mbox{\boldmath$\scriptstyle#1$}}
                     {\mbox{\boldmath$\scriptscriptstyle#1$}}}}
\fi